\documentclass[12pt]{article}
\usepackage{amsmath}
\usepackage{graphicx,psfrag,epsf}
\usepackage{enumerate}
\usepackage[left=1in,right=1in,top=1in,bottom=1in]{geometry}

\usepackage{setspace}
\usepackage{graphicx}
\usepackage{wrapfig}
\usepackage{epstopdf}
\usepackage{subcaption}
\usepackage{caption}
\usepackage{multirow}
\usepackage{nicefrac}
\usepackage{booktabs}
\usepackage{amsmath}
\usepackage{amsthm}
\usepackage[algo2e]{algorithm2e} 
\usepackage{verbatim} 
\usepackage{amssymb}
\usepackage{bbm}
\usepackage[english]{babel}
\usepackage{graphicx,epstopdf}
\RequirePackage[authoryear]{natbib}
\usepackage{hyperref}
\hypersetup{colorlinks=true,citecolor=blue}
\hypersetup{linkcolor=blue}
\usepackage{algorithm}
\usepackage[noend]{algpseudocode}
\usepackage{tikz}

\newcommand{\be}{\begin{eqnarray}}
\newcommand{\ee}{\end{eqnarray}}
\newcommand{\ba}{\begin{eqnarray*}}
\newcommand{\ea}{\end{eqnarray*}}
\newcommand{\bc}{\boldsymbol{c}}

\newcommand{\bphi}{\boldsymbol\phi}
\newcommand{\btheta}{\boldsymbol\theta}

\newtheorem{theorem}{Theorem}[section]

\newtheorem{lemma}[theorem]{Lemma}

\usepackage{bm}

\usepackage{array}
\newcolumntype{L}[1]{>{\raggedright\let\newline\\\arraybackslash\hspace{0pt}}m{#1}}
\newcolumntype{C}[1]{>{\centering\let\newline\\\arraybackslash\hspace{0pt}}m{#1}}
\newcolumntype{R}[1]{>{\raggedleft\let\newline\\\arraybackslash\hspace{0pt}}m{#1}}

\usepackage{array}

\begin{document}
\include{math-commands}
\begin{center}
    \LARGE{\bf A generalized hypothesis test for community structure in networks}\\
    \vspace{1 ex}
   \large{ \bf Eric Yanchenko and Srijan Sengupta}\\
   \large{ \bf North Carolina State University}
    
\end{center}

\begin{abstract}
\noindent
Researchers theorize that many real-world networks exhibit community structure where within-community edges are more likely than between-community edges. While numerous methods exist to cluster nodes into different communities, less work has addressed this question: given some network, does it exhibit {\it statistically meaningful} community structure? We answer this question in a principled manner by framing it as a statistical hypothesis test in terms of a general and model-agnostic community structure parameter. Leveraging this parameter, we propose a simple and interpretable test statistic used to formulate two separate hypothesis testing frameworks. The first is an asymptotic test against a baseline value of the parameter while the second tests against a baseline model using bootstrap-based thresholds. We prove theoretical properties of these tests and demonstrate how the proposed method yields rich insights into real-world data sets.
\end{abstract}



\newpage

\section{Introduction}\label{sec:intro}

Networks are everywhere in the modern world.
From social media \citep{kane2014s, guo2020online} to infrastructure \citep{mason2007graph} to epidemiology \citep{leitch2019toward}, many fields gather and analyze network data. The growth of the discipline of network science in the past two decades has brought along with it an interesting phenomenon: despite being observed in vastly different fields, many networks share similar structural properties \citep{fotouhi2019evolution}. One of the most common structural properties is {\it community structure}, or the occurrence of tightly-knit groups of nodes. Because communities can yield insights into node characteristics shared within these groups, various
\textit{community detection} methods have been proposed to assign nodes to communities, including spectral methods \citep{ng2002spectral, rohe11, jin::2015aa, sengupta2015spectral} and greedy algorithms \citep{Clauset:2004aa, blondel2008fast}. 
While the community detection problem has received substantial attention, less work has considered whether a given network demonstrates a statistically meaningful community structure. In other words, is the near-ubiquity of community structure in observed networks informative about the underlying data-generating mechanisms, or is it simply spurious false positive results due to random noise or some unrelated feature such as degree heterogeneity?

As a motivating example, consider the classical influence maximization (IM) task where the goal is to select a small number of seed nodes such that the spread of influence is maximized on the entire network \citep{kempe2003maximizing, yanchenko2023influence}. In networks lacking community structure, \cite{osawa2015selecting} showed that seeding nodes based on simple, centrality-based heuristics yields seed sets with substantial spread. On the other hand, when networks exhibit a community structure, selecting nodes from the same community results in ineffective seed sets due to nodes sharing many similar neighbors. In this case, more sophisticated seeding algorithms are needed. Thus, understanding the significance of the community structure in the network is paramount for adequate seed selection in the IM problem.

This work aims to develop a formal statistical hypothesis test for community structure. The framework of statistical hypothesis testing consists of four fundamental components: (1) the model parameter of interest, (2) a test statistic that is typically based on an estimator of the model parameter, (3) a null model that reflects the absence of the property of interest, and (4) a rejection region for the test statistic. While previous literature exists on this problem, there has been a minimal emphasis placed on ingredients (1) and (3). In particular, none of the existing work identifies an underlying model parameter and the null model has not been studied thoroughly. Two popular choices for nulls are the Erd\H{o}s-R\'{e}nyi (ER) \citep{bickel2016:aa, yuan2022testing} and configuration model \citep{Italy::2010aa, Palowitch:2018aa, li2020aa}. While these testing methods carry rigorous statistical guarantees,  choosing these null models means that they effectively test against the null hypothesis that the network is generated from a specific null model, rather than testing against the null hypothesis that there is no community structure. This leads to problems in empirical studies because the ER model, for example, is so unrealistic that almost all real-world networks would diverge from it, leading to many false positives. On the other hand, it may be possible for a configuration model to have a small amount of community structure itself. Thus, a careful treatment of the null model is essential for a statistical test to be relevant for applied network scientists.

The main contributions of this paper follow the four components of the hypothesis testing framework. From first-principles, we describe a model-agnostic parameter based on expected differences in edge densities which forms the basis of our statistical inference framework. Second, we propose an intuitive and interpretable test statistic which is directly connected to the model parameter. We leverage the model parameter and test statistic to formulate two types of hypothesis tests. The first is based on a user-specified threshold value of the parameter, which induces a model-agnostic test. For the second type, instead of specifying a baseline value of the parameter, the user specifies a baseline model or network property to test against. We derive theoretical results for the asymptotic cutoff in the first test and bootstrap cutoff in the second. Finally, we apply our method to well-studied real-world network datasets in the community structure literature. The results are insightful, as our method yields rich, new insights about the underlying network structure. Source code for this work is available on GitHub: \url{https://github.com/eyanchenko/NetHypTest}. 
The roadmap for the rest of this paper is as follows: in Section \ref{sec:method} we propose the model parameter and corresponding estimator, as well as present the first (asymptotic) hypothesis test. Section \ref{sec:3} discusses the baseline model test with a  bootstrap threshold. We apply the method to synthetic data in Section \ref{sec4} and real world datasets in Section \ref{sec:data}. We close by discussing the method in Section \ref{sec:disc}.

\section{Model parameter and baseline-value testing framework}\label{sec:method}
\subsection{Notation}
For this work,  we will only consider simple, unweighted and undirected networks with no self loops. 
Consider a network with $n$ nodes and let $A$ denote the $n\times n$ adjacency matrix where $A_{ij}=1$ if node $i$ and node $j$ have an edge, and 0 otherwise.
We write $A\sim P$ as shorthand for $A_{ij}|P_{ij}\sim\mathsf{Bernoulli}(P_{ij})$ for $1\leq i<j\leq n$ and define a community assignment to be a vector $\bc\in\{1,\dots,K\}^n$ such that $c_i=k$ means node $i$ is assigned to community $k\in\{1,\dots,K\}$. We also introduce the following notation: for scalar sequences $a_n$ and $b_n$, $a_n=O(b_n)$ means that $\lim_{n\to\infty} a_n/b_n\leq M$ for some constant $M$ (which could be 0) and $a_n=o(b_n)$ means that $\lim_{n\to\infty} a_n/b_n=0$. For a sequence of random variables, $X_n=O_p(Y_n)$ means that $X_n/Y_n\to_P M$ (which could be 0) and $X_n=o_p(Y_n)$ means that $X_n/Y_n\to_P 0$. 

\subsection{Expected Edge Density Difference (\textsf{E2D2}) parameter and estimator}\label{sec:prob}
The first step of the hypothesis test is to identify the model parameter. 
Since there is no universal metric to quantify community structure, we construct one from the first principles that applies to a large class of network models. 
For an {\it observed} network, a natural global measure of the strength of the community structure is the difference between the intra- and inter- community edge densities \citep[see page 83-84 of][]{Fortunato:2010aa}. The larger this difference, the more prominent the community structure is in the network. Now, this definition makes sense at the {\it sample} level for a realized network, but we seek the model parameter at the {\it population} level, or the parameter that generates this network. For this, we propose the {\it Expected Edge Density Difference} (\textsf{E2D2}) parameter. Consider a network model $P$ and let $\bc$ be a community assignment. Define
\begin{equation}
\begin{aligned}
    {\bar p_{\text{in}}}(\bc) 
    &= \frac{1}{\sum_{k=1}^K\binom{n_k}{2}}\sum_{i<j} P_{ij}\ \mathbbm{1}(c_i=c_j)  \ \ \ \text{ and }\ \ \ 
    \bar p_{\text{out}}(\bc) 
    &= \frac{1}{\sum_{k> l} n_k n_l}\sum_{i<j} P_{ij}\ \mathbbm{1}(c_i\neq c_j) ,
\end{aligned}
\end{equation}
where
$\mathbbm{1}(\cdot)$ is the indicator function.
Here, ${\bar p_{\text{in}}}(\bc)$ and $\bar p_{\text{out}}(\bc)$ are the average {\it expected} intra- and inter- community edge densities, respectively, hence the name {\it Expected} Edge Density Difference. 
This definition is sensible only when $1<K<n$ for if $K=1$, then $\bar p_{\text{out}}(\bc)$ is ill-defined and the same is true for $\bar p_{\text{in}}(\bc)$ if $K=n$. Intuitively, a data-generating mechanism with large ${\bar p_{\text{in}}}(\bc) - \bar p_{\text{out}}(\bc)$ is likely to produce a network with a large difference in observed intra- and inter- community edge density and, therefore, prominent community structure. This difference, however, should be adjusted with respect to the overall sparsity of the network and the number of groups each node can be assigned.
So, we propose the \textsf{E2D2} parameter $\gamma(\bc,P)$ as
\begin{equation}\label{eq:gamma}
    \gamma(\bc,P)
    :=\frac1K\frac{\bar p_{\text{in}}(\bc)-\bar p_{\text{out}}(\bc)}{\bar p},
\end{equation}
where $\bar p = \sum_{i>j} P_{ij}/\binom{n}{2}$ is the overall probability of an edge between nodes in the network. We also define

\begin{equation}
    \tilde\gamma(P)=\max_{\bc}\{\gamma(\bc,P)\}
\end{equation}
as the maximum value of the \textsf{E2D2} parameter where the maximization is taken over the candidate community assignments $\bc$. In the Supplemental Materials, we sketch a proof showing that $\bar p_{\text{in}}(\bc)-\bar p_{\text{out}}(\bc)\leq \bar p K$. Thus, the $K^{-1}$ term ensures that the \textsf{E2D2} parameter is always less than one. Indeed, $\gamma(\bc,P)=1$ if and only if $\bar p_{\text{out}}=0$ and there are $K$ equally-sized communities. By construction, this parameter has a natural connection to the intuitive notion of community structure since larger values correspond to more prominent levels of community structure in the data-generating process. 

One of the key advantages of the \textsf{E2D2} parameter is that it is general and model-agnostic. By {\it model-agnostic}, we mean that the definition applies to any network-generating model where edges are conditionally independent given the model parameters $P_{ij}$. This encompasses a wide-range of models including: Erd\H{o}s-R\'{e}nyi \citep{erdos1959}, Chung-Lu \citep{chung2002average}, stochastic block models \citep{holland1983stochastic}, degree corrected block models \citep{karrer2011stochastic}, latent space models \citep{hoff2002latent}, random dot product graphs \citep{athreya2017statistical} and more. Some other models such as the Barab{\'a}si-Albert \citep{barabasi1999emergence}, configuration \citep{fosdick2018configuring} and exponential random graph models \citep{robins2007introduction}, however, do not fit into this framework. While by no means applicable to \textit{all} network models, the general definition of the \textsf{E2D2} parameter, requiring only the conditional independence of edges, gives the parameter great flexibility.


The second ingredient in our hypothesis testing recipe is an estimator of the \textsf{E2D2} parameter. Using the same notation as above, define
\begin{equation}
\begin{aligned}
    {\hat p_{\text{in}}}(\bc) 
    &= \frac{1}{\sum_{k=1}^K\binom{n_k}{2}}\sum_{i<j} A_{ij}\ \mathbbm{1}(c_i=c_j) \ \ \ \text{ and }\ \ \ 
    \hat p_{\text{out}}(\bc) 
    &= \frac{1}{\sum_{k> l} n_k n_l}\sum_{i<j} A_{ij}\ \mathbbm{1}(c_i\neq c_j),
\end{aligned}
\end{equation}
Then we estimate $\gamma(\bc,P)$ from  (\ref{eq:gamma}) as
\begin{equation}
    T(\bc,A)
    :=\frac1K\frac{\hat p_{\text{in}}(\bc)-\hat p_{\text{out}}(\bc)}{\hat p},
\end{equation}
where $\hat p=\sum_{i>j}A_{ij}/{\binom{n}{2}}$ and $\hat p_{\text{in}}(\bc),\hat p_{\text{out}}(\bc)$ and $\hat p$ are the sample versions of $\bar p_{\text{in}}(\bc), \; \bar p_{\text{out}}(\bc)$, and $\bar p$, respectively. 
In other words, $T(\bc,A)$ is the {\it observed} edge density difference, the sample version of the \textsf{E2D2} parameter. 
Below we find the maximum of this test statistic over all possible community labels $\bc$ so we also introduce the notation
\begin{equation}\label{eq:stat}
    \tilde T(A)=\max_{\bc}\{T(\bc,A)\}.
\end{equation}
Since $\hat p$ depends on $A$ but not on $\bc$, $\tilde T(A)$ maximizes the intra-community edge probability over the candidate values of $\bc$ with a penalty for larger inter-community edge probability, akin to the objective function in \cite{mancoridis1998}.

This metric has a natural connection to the well-known Newman-Girvan {\it modularity} quantity \citep{Newman:2006aa}. The modularity, $Q(\bc,A)$, of a network partition $\bc$ is defined as
\begin{equation}\label{eq:mod}
    Q(\bc,A)
    =\frac1{m}\sum_{i<j}\left(A_{ij}-\frac{d_id_j}{2m}\right)\mathbbm{1}(c_i=c_j)
\end{equation}
where $d_i$ is the degree of node $i$ and $m$ is the total number of edges in the network. Rearranging (\ref{eq:mod}), we see that
\begin{equation}
    Q(\bc,A)
    =\frac1m\sum_{i<j}A_{ij}\mathbbm{1}(c_i=c_j) - \frac1{2m^2}\sum_{i<j} d_id_j\mathbbm{1}(c_i=c_j).
\end{equation}
The connection is now immediate: the first term is the (scaled) number of intra-community edges and has a one-to-one relationship with $\hat p_{\text{in}}(\bc)$. The second term can be thought of as the penalty term for the expected number of edges for a random network with given degree sequence. In light of these similarities, the proposed estimator has a key advantage compared to modularity. The penalty term for modularity assumes the configuration model as the null model, i.e., comparing the strength of the community structure against a random network with identical degree sequence. The penalty term for the proposed method $\hat p_{\text{out}}(\bc)$, however, is not model-dependent. Hence, any model could be chosen as the null model, giving the proposed estimator far greater flexibility than modularity. The numerator of $T(\bc,A)$ can also be written in the general modularity formulation of \cite{bickel2009nonparametric}.

\subsection{Algorithm for computing the \textsf{E2D2} estimator}\label{sec3}

The \textsf{E2D2} estimator is also of independent interest as an objective function for community detection. Finding the maximum of $T(\bc,A)$ is a combinatorial optimization problem with $O(K^n)$ solutions. Thus, an exhaustive search is clearly infeasible for even moderate $n$ so we propose a greedy, label-switching algorithm to approximate $\tilde T(A)$. We briefly explain the ideas here and present the full algorithm in Algorithm 1. First, each node is initialized with a community label $c_i\in\{1,\dots,K\}$. Then, for each node $i$, its community assignment is switched with all neighboring communities. The new label of node $i$ is whichever switch yielded the largest value of the \textsf{E2D2} estimator (or it is kept in the original community if none of the swaps increased $T(\bc,A)$). This process repeats for all $n$ nodes. The algorithm stops when all nodes have been cycled through and no labels have changed. The current labels, $\bc$, are then returned. The assumption that $K$ is known is rather strong and unrealistic for most real-world networks. We view it as reasonable here, however, since the goal of this work is not primarily to propose a new community detection algorithm. In practice, any off-the-shelf method can be used to estimate $K$ before using the proposed algorithm.

\begin{algorithm}[]
\SetAlgoLined
\KwResult{Community labels $\bc$}
 {\bf Input: } $n\times n$ adjacency matrix $A$, number of communities $K$\;
 
 Initialize labels $\bc\in\{1,\dots,K\}^n$\;
 
 $run = 1$\;
 
 \While{$run > 0$}{
 
 $run =0$\;
 
 Randomly order nodes\;
 
  \For{$i$ in $1,\dots,n$}{
  Find neighboring communities, $K_i$, of node $i$: $K_i=\{k_1,\dots,k_{|K_i|}\}$\;
  
  Swap label of node $i$ with all $k_j\in K_i$: $\bc_j^*=\bc$, $(\bc_j^*)_i=k_j$\;
  
  $\bc^*=\arg\max_{j}\{T(\bc^*_j,A)\}$\;
  
  \If{$T(\bc^*,A) > T(\bc,A)$}{
  
  $\bc\longleftarrow\bc^*$\;
  
  $run = 1$
 }
   
  }
 }
 \caption{Greedy}
 \label{alg:app}
\end{algorithm}

\subsection{Baseline-value test}

We now leverage the \textsf{E2D2} parameter and estimator to formulate our first hypothesis test. Because this parameter is interpretable and meaningful as a descriptor of the network-generating process, we consider the scenario where the researcher has a problem-specific benchmark value of the \textsf{E2D2} parameter that she would like to test against. In other words, the baseline value has domain-relevant meaning as ``no community structure." Then we must determine whether {\it any} assignment exceeds this threshold so the formal test is:
\begin{equation}\label{eq:h0math}
    H_0:  \tilde{\gamma}(P)\leq\gamma_0\ vs.\
    H_1:\tilde{\gamma}(P)>\gamma_0,
\end{equation}
for some $\gamma_0\in[0,1)$. Naturally, we reject $H_0$ if 
\begin{equation}\label{eq:max}
    \tilde T(A)=\max_{\bc} \{T(\bc, A)\}>C
\end{equation}
for some cutoff $C$ that depends on the network size $n$ and null value $\gamma_0$. 

Now, to obtain a test with level $\alpha$, we should set $C$ as the $(1-\alpha)$ quantile of the null distribution of $\tilde T(A)$. This depends on the data-generating matrix $P$ under the null hypothesis, and so implicitly also depends on $\tilde\gamma(P)=\gamma_0$.
But this is a difficult task since the test statistic is the maximum taken over $O(K^n)$ possible community assignments and these random variables are highly correlated. We propose to sidestep this difficult theoretical problem with an asymptotic cutoff. We first make the following three assumptions.\\ {\it A1. For any candidate community assignment, at least two community sizes must grow linearly with $n$.}\\ 
{\it A2. The number of communities $K_n$ is known, but is allowed to diverge.} \\
{\it A3. $\log^{1/2}K_n/(n^{1/2}\bar pK_n)\to0$ as $n\to\infty$.} \\
A1 lays down a basic requirement for any legitimate candidate community assignment, since otherwise, one community will dominate the entire network, while A2 shows that our theory holds even if the number of communities $K_n$ goes to infinity. A3 is a sparsity requirement for $\bar p$ that depends on the asymptotics of $K_n$. If $K_n\equiv K$ is fixed \citep[e.g.][]{bickel2009nonparametric, senguptapabm}, then we require $n^{1/2}\bar p\to\infty$. A typical sparsity assumption is that $\bar p=O(n^{-1})$ such that the expected number of edges in the network grows linearly with $n$. Our result requires a stronger condition which means we are in the {\it semi-dense} regime. While this is not ideal, proofs in the {\it dense} regime (fixed $p$) are common in the literature \citep[e.g.,][]{bickel2016:aa}. Our work, in fact, holds under less stringent conditions, i.e., $n^{1/2}\bar p\to\infty$ but allows $\bar p\to 0$. Indeed, this assumption implies that the test has larger power when the network is denser and/or has more communities since this leads to a smaller cutoff, as we see in the formal result that now follows.

\begin{theorem}\label{thm:ag}
    Let $A\sim P$ and consider testing $H_0:\tilde\gamma(P)\leq\gamma_0$ as in (\ref{eq:h0math}).  Let A1 and A2 be true and consider the cutoff
    \begin{equation}
        C = \left(\gamma_0+\frac{k_n}{K_n\hat p}\right)(1+\epsilon)
    \end{equation}
    where $k_n=\{(\log K_n)/n\}^{1/2}$ and arbitrarily small $\epsilon>0$ chosen by the user. Then when the null hypothesis is true $(\tilde\gamma(P)\leq\gamma_0)$, the type-I error goes to 0, i.e., for any $\eta>0$, 
    $$
        \lim_{n\to\infty}\mathsf{P}\{\tilde T(A) > C\mid H_0\}\leq \eta.
    $$
    If the alternative hypothesis is true $(\tilde\gamma(P)>\gamma_0)$, then the power goes to 1, i.e.,
    $$
        \lim_{n\to\infty}\mathsf{P}\{\tilde T(A) > C\mid H_1\} > 1- \eta.
    $$
\end{theorem}
A proof of the theorem, as well as proofs of all subsequent theoretical results, are left to the Supplemental Materials.
The proof approximates the cutoff under the null hypothesis using a union bound and then leverages Hoeffding's inequality to show that the probability of failing to reject under the alternative hypothesis goes to 0. The cutoff depends on $K$ and, for theoretical purposes, we assume that $K$ is known. In practice, we run a community detection algorithm on the network \citep[e.g., Fast Greedy algorithm of][]{Clauset:2004aa} and then use the number of communities returned by this algorithm, $\hat K$, to find $\tilde T(A)$ and construct the threshold. Additionally, A3 ensures that the $k_n/(K\hat p)$ term converges to 0 such that $C\to\gamma_0$ as $n\to\infty$. In practice, we confirm the necessity of this assumption as this test has larger power for denser networks (large $\hat p$) and more communities.

This test is fundamentally different from existing ones in the literature because the practitioner chooses the value of $\gamma_0$ for her particular problem, inducing a model-agnostic test. In other words, the null hypothesis is not a baseline model, but instead a baseline quantity of community structure in the network-generating process. Since the \textsf{E2D2} parameter has a natural connection to community structure, this test is also more easily interpretable with respect to this feature. Indeed, rejecting the null hypothesis means that the data-generating matrix for the observed network has greater community structure (as measured by the \textsf{E2D2} parameter) than some baseline value ($\gamma_0$).

A natural question that arises is how to choose $\gamma_0$. We stress that this choice depends on the domain and question of interest. There are some special cases, however, that yield insights into selecting a meaningful value. For example, assume that the practitioner believes that her network has two roughly equal-sized communities. Then setting $\gamma_0=(\mu-1)/(\mu+1)$ is equivalent to testing whether the average intra-community edge density is more than $\mu>1$ times larger than the average inter-community edge density, i.e., $\bar p_{\text{in}} \geq \mu\bar p_{\text{out}}$.

The special case of $\gamma_0=0$ is also worthy of further discussion. It is trivial to show that if $P$ is from an Erd\H{o}s-R\'{e}nyi (ER) model \citep{erdos1959} where $P_{ij}=p$ for all $i,j$, then $\tilde\gamma(P)=0$. We show in the Supplementary Materials, however, that the converse of this statement is also true, i.e., $\tilde\gamma(P)=0$ {\it only if} $P$ is from an ER model. This means that setting $\gamma_0=0$ is equivalent to testing against the null hypothesis that the network is generated from an ER model. In other words, any other network model will reject this test when $\gamma_0=0$. But there are many models (e.g., Chung-Lu \citep{chung2002average}, small world \citep{watts1998collective}) which may not be ER but also do not intuitively have community structure. This connection between the model-agnostic \textsf{E2D2} parameter and its behavior under certain model assumptions motivates the test in the following section.

\section{Baseline-model test}\label{sec:3}
In the previous section, we derived a hypothesis test based on a user-defined benchmark value that does not refer to a null model. There may be situations, however, where the practitioner does not have a meaningful way to set the null parameter $\gamma_0$. In this case, we set the null hypothesis in reference to a particular null model and/or model property. If $P(\bphi)$ is the true data-generating model for $A$ defined by the parameters $\bphi$, then instead of testing $\tilde\gamma(P(\bphi))\leq \gamma_0$, the null hypothesis is now that $\tilde\gamma(P(\bphi))$ is less than or equal to the largest value of the \textsf{E2D2} parameter under the null model. Since the specific set of parameters for the null model is typically unknown, they are estimated from the observed network as $\hat{\bphi}$. To fix ideas, we provide the following two examples.


\paragraph{Erd\H{o}s-R\'{e}nyi model:}
The simplest null model to consider is the ER model where $P_{ij}=p$ for all $i,j$. Thus, the value of $p$ completely defines an ER model. We estimate $p$ by taking the average edge probability of the network, i.e., $\hat p=\sum_{i<j}A_{ij}/\{n(n-1)/2\}$. Then this test determines whether the observed value of the \textsf{E2D2} estimator is greater than what could arise if the network was generated from an ER model. Recall that this is equivalent to the test in (\ref{eq:h0math}) setting $\gamma_0=0$.


\paragraph{Chung-Lu model:}
Another sensible null model to consider is the Chung-Lu (CL) model \citep{chung2002average} where $P_{ij}=\theta_i\theta_j$ for some weight vector $\boldsymbol\theta=(\theta_1,\dots,\theta_n)$ which uniquely defines the model. This model is similar to the configuration model except instead of preserving the exact degree sequence, it preserves the expected degree sequence. We estimate $\btheta$ using the rank 1 Adjacency Spectral Embedding \citep{sussman2012consistent}:
\begin{equation}\label{eq:ase}
    \hat{\btheta}
    =|\hat\lambda|^{1/2}\hat{\bf u}
\end{equation}
where $\hat\lambda$ is the largest-magnitude eigenvalue of $A$ and $\hat{\bf u}$ is the corresponding eigenvector. Now the test is whether the value of the \textsf{E2D2} estimator is greater than what likely would have been observed if the CL model generated the observed network.


\subsection{Bootstrap test}

To carry out this test, we propose a bootstrap procedure. We describe the method for the CL null but full details for the ER and CL null can be found in Algorithm \ref{alg:cl}. Additionally, the bootstrap test can be trivially modified to test against many other null models.

This approach directly estimates the $1-\alpha$ quantile of the null distribution of $\tilde T(A)$ with a parametric bootstrap and then uses this quantity as the testing threshold. In particular, we first compute the test statistic $\tilde T(A)$ as in (\ref{eq:stat}). Since the null distribution of $\tilde T(A)$ is unknown, we must simulate draws from this distribution in order to have a comparison with our observed test statistic.  So we next estimate $\boldsymbol\theta$ with the adjacency spectral embedding (ASE) as in (\ref{eq:ase}). 
Then, for $b=1,\dots,B$, we draw $\hat{\boldsymbol\theta}^*_b=(\hat\theta_{b1}^*,\dots,\hat\theta_{bn}^*)^T$ with replacement from $\hat{\boldsymbol\theta}=(\hat\theta_1,\dots,\hat\theta_n)^T$, generate a CL network $A_b^*$ with $\hat{\boldsymbol\theta}^*_b$ and find $\tilde T_b^*=\max_{\bc}\{T(A_b^*, \bc\}$. The empirical distribution of $\{\tilde T^*_b\}_{b=1}^B$ serves as a proxy for the null distribution of $\tilde T(A)$ so the $p$-value is
\begin{equation}
    p\text{-value}=\frac1B\sum_{b=1}^B\mathbbm{1}\{\tilde T_b^* \geq \tilde T(A)\}
\end{equation}
and we reject $H_0$ if the $p$-value is less than a pre-specified $\alpha$. 

\begin{algorithm}[]
\SetAlgoLined
\KwResult{$p$-value}
 {\bf Input: } $n\times n$ adjacency matrix $A$, number of iterations $B$, null model $\mathcal M$\;
 
  Compute $\tilde T(A) = \max_{\bc}\{T(A,\bc)\}$ as in (\ref{eq:stat}) \;
  
  \If{$\mathcal M$=ER} {Compute $\hat p=\sum_{i<j}A_{ij}/\{n(n-1)/2\}$\;}
  \If{$\mathcal M$=CL} {Compute $\hat{\boldsymbol\theta} = \hat\lambda^{1/2}\hat {\bf u}$ as in (\ref{eq:ase}) \;}

 \For{B times}{
   \If{$\mathcal M$=ER} {$A^*_b\leftarrow$ ER network with $\hat p$\;}
  \If{$\mathcal M$=CL} {Draw $\hat{\boldsymbol\theta}^*_b=(\hat\theta_{b1}^*,\dots,\hat\theta_{bn}^*)^T$ with replacement from $\hat{\boldsymbol\theta}=(\hat\theta_1,\dots,\hat\theta_n)^T$\; (\ref{eq:ase}) \;
  
  $ A^*_b\leftarrow$ CL network with $\hat {\boldsymbol{\theta}}^*_b$\;}
 
   Compute $ \tilde T^*_b= \max_{\bc}\{T(A_b^*, \bc)\}$ \;
   
  }
  $p$-val$=\sum_b\mathbbm{1}\{\tilde T^*_b \geq \tilde T(A)\}/B$
 
 \caption{Bootstrap hypothesis test}
 \label{alg:cl}
\end{algorithm}

We highlight a nuanced but important distinction between the interpretation of the baseline-value and the baseline-model test. For each bootstrap iteration, we re-estimate the number of groups $K_b^*$ that is then used when computing $\tilde T_b^*=\max_{\bc}\{T(A^*_b,\bc)\}$. Therefore, it is possible that for a given iteration, $K^*_b\neq K$, where $K$ was estimated from the original network $A$. Thus, the results of the bootstrap test are {\it unconditional} on the value of $K$. Rejecting against the ER null, for example, means that the observed network has greater community structure than could be expected from a network generated from an ER model, and for any $K$ used to compute $\tilde T(A)$. Conversely, the baseline-value test results are {\it conditional} on $K$. Therefore, rejecting this test means that the observed network has stronger community structure than a network model with $\tilde\gamma(P)=\gamma_0$ could generate with $K$ groups.

Now, the bootstrap test naturally fits into the existing community detection testing literature as it considers a specific null model. Indeed, to the knowledge of the authors, this is the first\footnote{\cite{mukherjee2021testing} test for the significance of {\it degrees} in a DCBM, i.e., $H_0$: SBM vs.~$H_1$: DCBM whereas the proposed test considers $H_0:$ CL vs.~$H_1:$ DCBM.} statistical hypothesis test for the significance of communities in a degree corrected block model \citep{karrer2011stochastic} as the CL model serves as the null model of ``no communities.'' Nevertheless, the bootstrap method's strength is its flexibility, owing in part to the general definition of the \textsf{E2D2} metric, in that it can easily accommodate any null model\footnote{Subject to the modeling constraints mentioned in Section \ref{sec:prob}.}. Additionally, it allows for more general inference as multiple, realistic null distributions can be tested against, leading to a rich understanding of the network's community structure. The bootstrap test also yields an insightful visual tool where the observed value of the test statistic is plotted next to the bootstrap histogram. This tool helps in visualizing how the strength of community structure observed in the network compares to benchmark networks with the same density, or with the same degree distribution, and so on. We study these plots more in Section \ref{sec:data}. Lastly, a fundamental challenge of bootstrapping networks is that, in general, we only observe a single network. If we knew the model parameters (e.g., $p$ or $\boldsymbol\theta$) then it would be trivial to generate bootstrap replicates. 
Since the true model parameters are unknown, we must first estimate them and then generate networks using the estimated parameters. Thus, the quality of the bootstrap procedure depends on the quality of these estimates. In the following sub-section, we formally prove certain properties of this procedure.

\subsection{Bootstrap theory}
We now turn our attention to theoretical properties of the bootstrap. We want to show that, if $A,H\sim P(\bphi)$ and $\hat A^*\sim P(\hat{\bphi})$ where $\hat{\bphi}$ estimates $\bphi$ using $A$, then $\tilde T(\hat A^*)$ converges to $\tilde T(H)$. To have any hope of showing this result, $\hat A^*$ must be similar to $A$. We consider the {\it Wasserstein p-distance} and adopt the notation of \citep{levin2019bootstrapping}. Let $p\geq1$ and let $A_1,A_2$ be adjacency matrices on $n$ nodes. Let $\Gamma(A_1,A_2)$ be the set of all couplings of $A_1$ and $A_2$. Then the Wasserstein $p$-distance between $A_1$ and $A_2$ is
\begin{equation}
    W_p^p(A_1,A_2)
    =\inf_{\nu\in\Gamma(A_1,A_2)}\int d^p_{GM}(A_1,A_2)d\nu
\end{equation}
where 
\begin{equation}
    d_{GM}(A_1,A_2)
    =\min_{Q\in\Pi_n}{{n\choose 2}}^{-1}\frac12 \|A_1-QA_2Q'\|_1
\end{equation}
where $\Pi_n$ is the set of all $n\times n$ permutation matrices and $\|A\|_1=\sum_{i,j}|A_{ij}|$. The following results show that $\hat A^*$ converges in distribution to $A$ in the Wasserstein $p$-distance sense for both the ER and CL null. For all results in this section, assume that model parameters do not depend on $n$, i.e., $p_n=p\not\to0$.

\begin{lemma}\label{lem:1}
Let $A,H\sim ER(p)$ and $\hat A^*\sim ER(\hat p)$ where $\hat p=\sum_{i,j}A_{ij}/\{n(n-1)\}$. Then
$$
    W_p^p(\hat A^*,H)
    = O(n^{-1}).
$$
\end{lemma}
\begin{lemma}\label{lem:2}
Let $A,H\sim CL(\btheta)$ and $\hat A^*\sim CL(\hat{\btheta})$ where $\hat{\btheta}$ is found using (\ref{eq:ase}). Then
$$
    W_p^p(\hat A^*,H)
    = O(n^{-1/2}\log n).
$$
\end{lemma}
\noindent
Both proofs are based on Theorem 5 in \cite{levin2019bootstrapping}. These results show that for large $n$, the networks generated from the bootstrap model are similar to the networks generated from the original model. Since there is only one parameter to estimate in the ER model but $n$ in the CL model, it is sensible that the rate of convergence is much faster for the former.

Next, we show that for a fixed $\bc$, the distribution of the bootstrapped test statistic converges to the same distribution as the original test statistic. We only show this result for the ER null. First, we introduce some useful notation. Let $t(A,\bc)$ be the numerator of the \textsf{E2D2} test estimator, i.e.,
$$
    t(A,\bc)
    =\frac1{m_{\text{in}}}\sum_{i<c} A_{ij}\mathbbm{1}(c_i=c_j)- \frac1{m_{\text{out}}}\sum_{i<c} A_{ij}\mathbbm{1}(c_i\neq c_j) 
    :=\sum_{i<j}C_{ij}A_{ij}
$$
where $m_{\text{in}}$ and $m_{\text{out}}$ are the total possible number of intra- and inter- community edges, respecitvely, and $C_{ij}=m_{\text{in}}^{-1}$ if $c_i=c_j$ and $m_{\text{out}}^{-1}$ otherwise. Let $C_1=\sum_{i<j}C_{ij}$ and $C_2=\sum_{i<j}C_{ij}^2$. Additionally, let $K$ be the number of groups that a node can be assigned, which is assumed to be fixed\footnote{Note that for the ER data-generating model, there is effectively only one community. During the inference procedure, however, $K$ is unknown and must be estimated from the network. Thus, in this situation, $K$ is better considered as the number of groups a node can be assigned to, rather than the number of communities.}. Then we have the following result.

\begin{lemma}\label{lem:3}
Let $A,H\sim ER(p)$ and $\hat A^*\sim ER(\hat p)$ where $\hat p=\sum_{i<j}A_{ij}/\{n(n-1)/2\}$ and consider a fixed $\bc$. Furthermore, let $s_n^2=p(1-p)C_2$. Then
$$
    \frac1{s_n}\{T(H,\bc)-\gamma(H,\bc)\}\stackrel{d}{\to}\mathsf{N}(0,K^2p^2)
$$
and
$$
    \frac1{s_n}\{T(\hat A^*,\bc)-\gamma(H,\bc)\}\stackrel{d}{\to}\mathsf{N}(0,K^2p^2)
$$
\end{lemma}
\noindent
The proof is a simple application of the non-identically distributed central limit theorem and iterated expectations. This lemma implies that the \textsf{E2D2} estimator $T(A,\bc)$ consistently estimates the \textsf{E2D2} model parameter $\gamma(A,\bc)$ for a particular $\bc$. Additionally, the distribution of the test statistic converges to the same normal distribution, whether the network was generated from the original model or the bootstrap model. This result is more difficult to show for the CL null model. We cannot use the ideas from the proof of Lemma \ref{lem:3} because $\hat{\btheta}$ has a more complicated form and the bootstrap step is more involved than that of the ER null; nor can we use the results in \cite{levin2019bootstrapping} because the \textsf{E2D2} estimator cannot be written as $U$-statistic. 

Ideally, we would like to show that this result also holds when using the community assignment which maximizes the \textsf{E2D2} estimator. Unfortunately, showing this convergence for arbitrary statistics is difficult \citep[e.g.,][]{levin2019bootstrapping}. This is challenging in our particular case for several reasons. First, the \textsf{E2D2} estimator $\tilde T(A)$ is the maximum of $O(e^n)$ statistics $T(A,\bc_i)$, meaning the maximum is taken over a set of random variables which goes to infinity. Additionally, these variables are non-trivially correlated since they depend on the same adjacency matrix. Another angle to view the difficulty of this problem is that, for $\bc^*=\arg\max_{\bc}\{T(A,\bc\}$, $C_{ij}^*$ is now dependent on $A_{ij}$. Even computing the mean and variance of this estimator becomes difficult. Bootstrap theoretical results for the maximum of the test statistic is an important avenue for future work.

\section{Hypothesis Testing Simulations}\label{sec4}

\subsection{Settings}

We now study the performance of the proposed method on synthetic data. Our primary metric of interest is the rejection rate of the test under different settings. We consider two settings for the baseline-value test as well as settings for the baseline-model test with both the ER and CL nulls. In each setting, we first fix $\tilde\gamma(P)$ and increase the number of nodes $n$. Then we fix $n$ and increase $\tilde\gamma(P)$ and, in both cases, we expect an increasing rejection rate. We run 100 Monte Carlo simulations and compute the fraction of rejections. Our bootstrap method uses $B=200$ bootstrap samples and we fix the level of the test at $\alpha=0.05$. We chose the two Spectral methods proposed in \cite{bickel2016:aa} as benchmarks since these are leading and well-established methods with formal guarantees. Even though the authors suggest only using the adjusted method, we will still compare both since, similar to our proposed framework, the authors propose a version of the test with an asymptotic threshold and an adjusted version of the test with a bootstrap correction. 

\subsection{Test against baseline value}
First, we consider the baseline-value test using Theorem \ref{thm:ag}, i.e., we reject $H_0$ if
$$
    \tilde T(A)
    > \left(\gamma_0 + \frac{k_n}{K\hat p}\right)(1+\epsilon)
$$
where $k_n=\{(\log K)/n\}^{1/2}$ and $\epsilon=0.0001$.
We let $n=1000,1200,\dots, 3000$ and generate networks with $K$ communities. When $n=1000$, we let $K=2$ where 60\% and 40\% of the nodes are in each community, respectively. When $1000<n\leq 2000$, $K=4$ with 40\%, 20\%, 20\% and 20\% of the nodes distributed in each community. For $n>2000$, we have $K=6$ with 25\%, 15\%, \dots, 15\% of nodes in each community. The edge probabilities $P_{ij}$ are distributed such that
$$
    P_{ij}\stackrel{\text{indep.}}{\sim} \mathbbm{1}(c^*_i=c^*_j)\mathsf{Uniform}(p_l, p_u) + \mathbbm{1}(c^*_i\neq c^*_j)\mathsf{Uniform}(0.05,0.07)
$$
where $\bc^*$ corresponds to the true community labels. We set $(p_l,p_u)=(0.075,0.095)$, $(0.100, 0.120)$, and $(0.125, 0.145)$ when $K=2,4$ and $6$, respectively, in order to fix $\mathsf{E}\{\tilde\gamma(P)\}\approx 0.20$ for all parameter combinations. This data-generating model has a block structure but with heterogeneous edge probabilities, which means $P$ is not an SBM. We test against the null hypothesis $H_0: \tilde\gamma(P)\leq \gamma_0=0.10$. The results are in Figure \ref{fig:sims}(a). Since $\mathsf{E}\{\tilde\gamma(P)\}>\gamma_0$, the test should reject. The cutoff is asymptotic, however, so the test has low power for small $n$. But when $n> 2000$, the test has a high power as $n$ is large enough for the asymptotic results to apply.

For the next setting, we fix $n=5000$ and let $K=4$ with 40\%, 20\%, \dots, 20\% of the nodes in each community. The edge probabilities $P_{ij}$ are now distributed such that
$$
    P_{ij}\stackrel{\text{indep.}}{\sim} \mathbbm{1}(c^*_i=c^*_j)\mathsf{Uniform}(0.125, 0.175) + \mathbbm{1}(c^*_i\neq c^*_j)\mathsf{Uniform}(a,0.125)
$$
where $a=0.10, 0.09,\dots,0.01$. We find that $\mathsf{E}\{\tilde\gamma(P)\}=0.08, 0.09, \dots, 0.23$ for different values of $a$ and test against the null hypothesis $H_0: \tilde\gamma(P)\leq \gamma_0=0.10$. The results are in Figure \ref{fig:sims}(b). When $a=0.10$, $\tilde\gamma(P)=0.08<\gamma_0=0.10$ so we would expect the test to fail to reject which it does. The test should have a high rejection rate when $a=0.07$ since $\tilde\gamma(P)=0.12>\gamma_0$. The power of the test, however, does not increase to one until $\tilde\gamma(P)=0.15$. This small discrepancy is due to the fact that it is an asymptotic cut off and we are generating networks with a finite number of nodes. When $\tilde\gamma(P)\geq0.15$, the test consistently rejects as expected.

\subsection{Test against ER null}\label{sec:sim_bootvasym}

Next, we study the baseline-model test using the bootstrap procedure described in Algorithm \ref{alg:cl}. First we test against the null hypothesis that the network was generated from the ER model. We showed that this test is also equivalent to the asymptotic test with $\gamma_0=0$ so we can also compare the rejection threshold from Theorem \ref{thm:ag}. A natural alternative model to the ER is the stochastic block model (SBM) \citep{holland1983stochastic} where $P_{ij}=B_{c_i,c_j}$ and $\bc$ corresponds to the true community labels.  We let $n=250,500,1000,1500,2000$ and generate networks from an SBM with $K$ communities. For $n<1000$, we set $K=2$ with 60\% and 40\% of the nodes in each community. $K=4$ with a 40\%, 20\%,\dots,20\% split for $1000\leq n<2000$, and $K=6$ for $n=2000$ with 25\%, 10\%,\dots,10\% of nodes in each community. We fix the inter-community edge probability $B_{ij}=0.025$ for $i\neq j$ and set $B_{ii}=0.05, 0.075, 0.10$ when $K=2,4,6$, respectively, for $i=1,\dots,K$. This ensures a fixed $\tilde\gamma(P)\approx 0.33$. The results are in Figure \ref{fig:sims} (c). Since the network is generated from an SBM and we are comparing with an ER null, we expect the test to yield a large rejection rate. We can see that both Spectral methods and the proposed bootstrap approach having an increasing rejection rate with increasing $n$ and where the Spectral methods have a larger power. The asymptotic method has a large rejection rate for $n=250$ but then drops to zero for $n=500$ before increasing again at $n=1000$. The reasons for this is that for $n=250$, the number of communities $K$ is being overestimated. This causes the test statistic to inflate more than the cutoff, leading to a large rejection rate. For $n\geq 500$, $K$ is more accurately estimated so we see trends that are expected.

In the second scenario, we fix $n=1000$. Now, the intra-community edge probability is $B_{11}=B_{22}=0.05$ and inter-community edge probability is $B_{12}=0.05, 0.04,\dots, 0.01$. This means that $\tilde\gamma(P)=0$, $0.11$, $0.25$, $0.42$, $0.65$. The results are in Figure \ref{fig:sims} (d). When $\tilde\gamma(P)=0$, the network is generated from an ER model meaning the null hypothesis is true so we expect a low rejection rate. The unadjusted Spectral method has large Type I error and the bootstrap method rejects slightly more than the $\alpha$ level of the test. 
When $\tilde\gamma(P)>0$, the null hypothesis is false so we expect a large rejection rate. Both Spectral methods reach a power of one by $\tilde\gamma(P)=0.25$ while the bootstrap method does not reach this power until $\tilde\gamma(P)=0.42$. The asymptotic test, as expected, is the most conservative test. While the Spectral methods outperform the bootstrap test in both scenarios, this is to be expected because these methods were designed for this exact scenario and null. The proposed test is more general so it can be applied to many more settings, but is unlikely to beat a method designed for a specific scenario.

\subsection{Test against CL null}\label{sec:sim_cl}
\begin{figure}
    \centering
    \includegraphics[width=\textwidth]{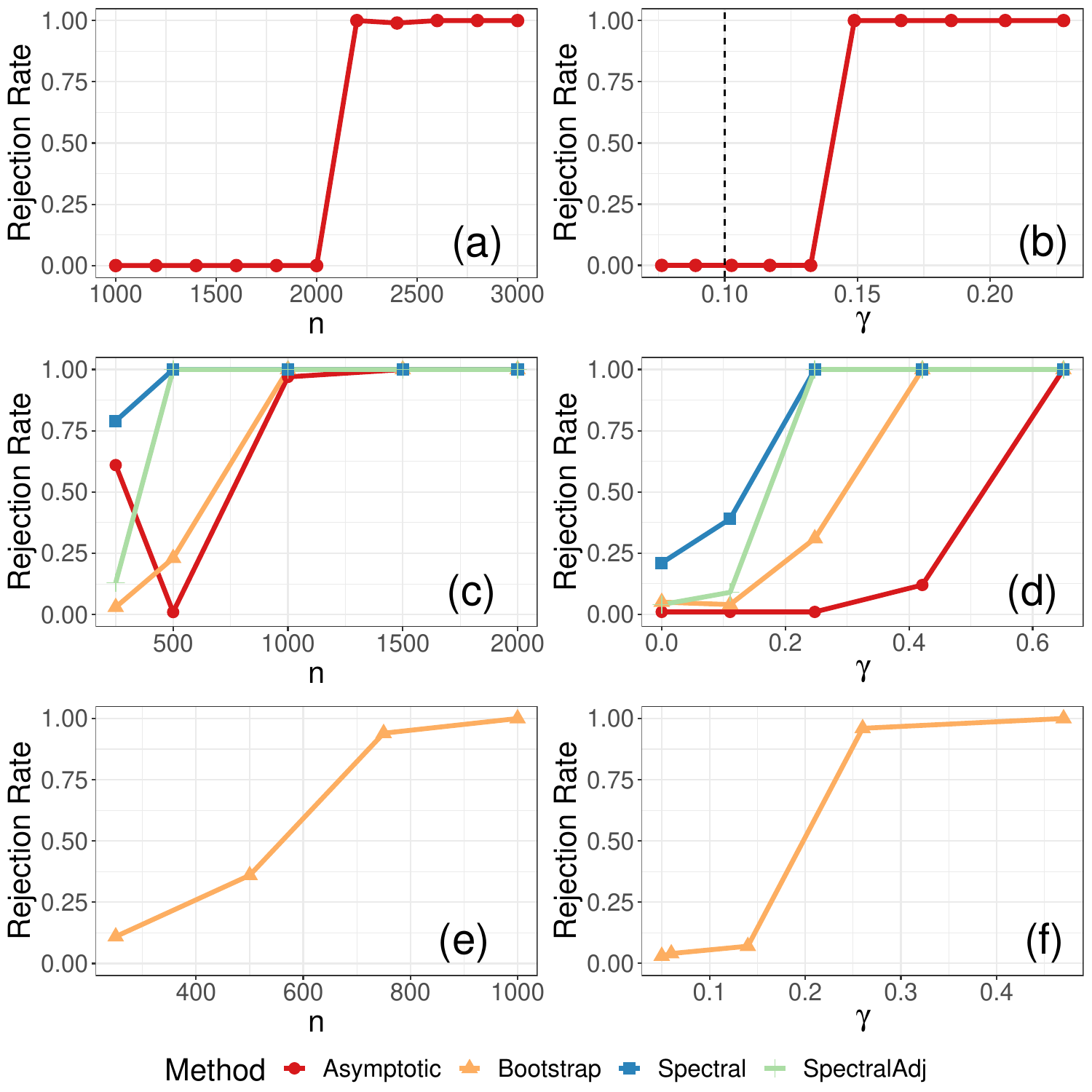}
    \caption{Rejection rates from simulation study. See Section \ref{sec4} for complete details. (a) baseline-value null with fixed $\tilde\gamma(P)$; (b) baseline-value null with fixed $n$; (c) Erd\H{o}s-R\'{e}nyi null with fixed $\tilde\gamma(P)$; (d) Erd\H{o}s-R\'{e}nyi null with fixed $n$; (e) Chung-Lu null with fixed $\tilde\gamma(P)$; (f) Chung-Lu null with fixed $n$}
    \label{fig:sims}
\end{figure}

Lastly, we study the baseline-model test where now the CL model functions as the null, again using Algorithm \ref{alg:cl}. We drop both Spectral methods for these simulations as the ER null is hard-coded into them, thus making these methods inapplicable for testing against the CL null. We also drop the asymptotic test from the comparison because it is unclear how to set $\gamma_0$ for this null model. In fact, finding a closed-form expression for the rejection threshold for the CL null is an interesting avenue of future work. For the alternative model, we consider the degree corrected block model (DCBM) \citep{karrer2011stochastic} where $P_{ij}=\theta_i\theta_j B_{c_i,c_j}$ and $\theta_i$ are node specific degree parameters. We generate networks from a DCBM with $n=250,500,\dots,1000$. For $n\leq 500$, we let $K=2$ where 60\% and 40\% of the nodes are in each community. For $n>500$, $K=3$ with 40\%, 30\%, and 30\% of the nodes distributed in each community. The degree parameters are generated $\theta_i\stackrel{\text{iid}}{\sim}\mathsf{Uniform}(0.2,0.3)$ and $B_{ii}=1$ for $i=1,\dots,K$. In order to preserve $\tilde\gamma(P)\approx 0.33$, $B_{ij}=0.5, 0.4$ when $K=2,3$, respectively, for $i\neq j$. The results are in Figure \ref{fig:sims}(e). Since the networks are generated from a DCBM, we expect a large rejection rate. We see that that rejection rate increases monotonically with $n$, reaching a power of one by $n=1000$.

For the second scenario, we fix $K=4$ with 40\%, 20\%, \dots, 20\% of the nodes in each community and $n=1000$. Let $B_{ii}=1$, $B_{ij}=1$, $0.8$, $\dots$, $0.2$ for $i\neq j=1,\dots,4$. Thus, $\tilde\gamma(P)=0.05, 0.06, 0.14, 0.26, 0.47$. The results are in Figure \ref{fig:sims}(f). When $B_{12}=1$ $(\tilde\gamma(P)=0.03)$, the networks are generated from the CL (null) model so we expect a low rejection rate and the bootstrap test has a low Type I error. When $B_{12}<1$ $(\tilde\gamma(P)>0.05)$, the null hypothesis is false so we expect a large rejection rate. The bootstrap reaches a power of one by $\tilde\gamma(P)=0.47$.

\section{Real data analysis}\label{sec:data}
We now study the proposed method on two networks: DBLP and hospital interactions. The DBLP is a computer science bibliography website and this network was extracted by \cite{gao2009graph} and \cite{ji2010graph}. Here, each node represents an author and an edge signifies that the two authors attended the same conference. Additionally, we only consider two author's research areas, databases and information retrieval. The hospital network \citep{Vanhems:2013} captures the interacts between patients and healthcare providers at a hospital in France. Each person is represented by a node and an edge signifies that they were in close proximity. 

\begin{figure}
\centering
\subfloat[DBLP]{\includegraphics[width = 3in]{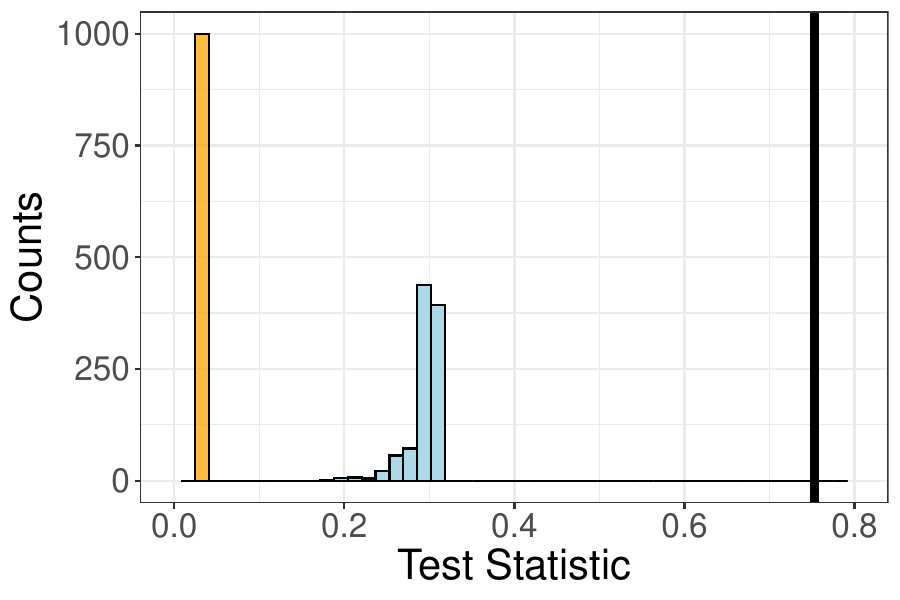}}
\subfloat[Hospital Encounters]{\includegraphics[width = 3in]{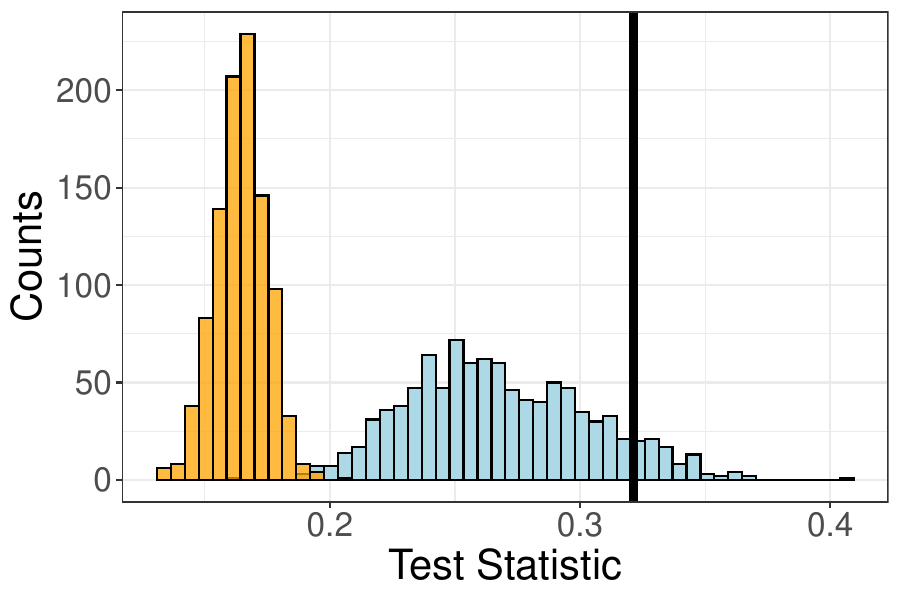}}
\caption{Histograms of bootstrap samples from the proposed method for the two real data sets. The orange histogram is with the Erd\H{o}s-R\'{e}nyi null, and the blue histogram is with the Chung-Lu null.
The vertical line (black) indicates the value of the test statistic.}
\label{fig:data}
\end{figure}

For each data set, we consider several metrics. First, we compute $\tilde T(A)$ and find the largest value of $\gamma_0$ such that the test in (\ref{eq:h0math}) is still rejected. We also compute the $p$-value and find the bootstrap histogram of the test statistic for the ER and CL null models (with $B=1,000$). Considering both null hypotheses together allows us to gain a richer understanding of the network. We compare the proposed method to the $p$-value from the adjusted Spectral method \citep{bickel2016:aa}. See Table \ref{tab:data} for numeric results and Figure \ref{fig:data} for histograms from the bootstrap method.
While Table \ref{tab:data} provides a succinct summary, the plots in Figure \ref{fig:data} provide more details and insights.
In these plots, the observed test statistic computed from the dataset is plotted as a vertical line along with histograms representing bootstrap distributions from various benchmark models.
These simple but informative plots give practitioners a reference of how the observed community structure compares to the range of community structure in various benchmarks.

\begin{table}[]
  \centering
    \begin{tabular}{lrr|cc|r|rr}
    & & & & &  {Spectral Adj.} & \multicolumn{2}{c}{Bootstrap}\\
    Network & $n$ & $m$ & $\tilde T(A)$& $\gamma_0$ & ER & ER & CL\\\hline
    DBLP & 2,203  &1,148,044  & 0.75  & 0.73   & $<0.001$ & 0.000 & 0.000 \\
    Hospital  & 75 & 1,139 & 0.32 & 0.22 & $<0.001$ & 0.000 & 0.104\\
    \end{tabular}
    \caption{The number of nodes $n$ and edges $m$ for real-world networks. $\tilde T(A)$ is the observed value of the \textsf{E2D2} parameter and $\gamma_0$ is the largest null value such that the baseline-value test would be rejected. Additionally, we report the $p$-values for the adjusted Spectral method and Bootstrap method against different null hypotheses (ER=Erd\H{o}s-R\'{e}nyi, CL=Chung-Lu).} 
    \label{tab:data}
\end{table}

For the DBLP network, since we only selected two research areas, we set $K=2$ to find $\tilde T(A)$. The observed value of $\tilde T(A)=0.75$ is quite large and, due to the network's size and density, would reject the base-line value test up to $\gamma_0=0.73$. This means that if $P$ generated this network, then we can assert that $\tilde\gamma(P)\geq 0.73$. Additionally, all $p$-values for the model-based tests are effectively zero and, moreover, the observed test statistic is in the far right tail of both bootstrap null distributions. Since both tests are rejected, then it is very unlikely that the network was generated from an ER or CL model and, instead, implies that there is highly significant community structure in this network. This finding accords with existing literature \citep[e.g.,][]{senguptapabm}.

For the Hospital network, the test of $H_0:\gamma_0=0$ is rejected, as are both tests with ER null. This is a sensible result since we showed that $\gamma_0=0$ is equivalent to testing against the ER null. The $p$-value for the CL null, however, is not significant at the $\alpha=0.05$ level, meaning that this network does not have more community structure than we would expect to occur from a CL network by chance. So while there is strong evidence that the network diverges from an ER model, these findings indicate that perhaps the low ER-null $p$-values are due to degree heterogeneity rather than community structure. Indeed, the histogram shows how the test statistic is very unlikely to have been drawn from the ER distribution, but is reasonably likely to have come from the CL distribution since this distribution has a greater mean and variance. Using an ER null alone would have led to the conclusion that there is community structure in this network. By using multiple nulls together with the proposed method, however, we gain a fuller understanding of the network by concluding that degree heterogeneity may be masquerading as community structure.

\section{Discussion}\label{sec:disc}

In this work, we proposed two methods to test for community structure in networks. These tests are rooted in a formal and general definition of the \textsf{E2D2} parameter. This metric is simple, flexible, and well-connected to the conceptual notion of community structure which we argue makes it a more principled approach. In fact, the test statistic can even be used as a descriptive statistic to quantify the strength of community structure in a network. Existing methods are based on specific random graph models, such as the ER model, which are implicitly presumed to be the only models that do not have community structure. While our second testing approach fits into this framework as well, the general nature of the \textsf{E2D2} parameter means that we can test against nearly any null model (subject to conditional edge independence) to obtain a richer set of practical insights compared to existing methods. Given a network, we recommend that practitioners first carry out the test against the ER null.  If this test is rejected, further tests should be carried out to check whether the it could be due to some other network feature like degree heterogeneity. Thus, the method not only helps decide whether the network appears to exhibit community structure but also helps understand the source of this ostensible community structure. 

There are several interesting future research directions. First, the proposed \textsf{E2D2} parameter and bootstrap testing framework could be adapted for sequential testing. In \cite{ghosh2013selecting}, the authors propose a general framework for sequentially testing for $H_0:K$ vs.~ $H_1:K+1$ communities in the network. Using the proposed parameter, we would first find $\tilde T(A)$ setting $K=2$, and then generate bootstrap samples with $K=1$ (ER model). If this test is rejected, we find $\tilde T(A)$ with $K=3$ and generate bootstrap samples from an SBM with $K=2$. Note that scaling the \textsf{E2D2} parameter by $K$ ensures a fair comparison of the metric across different values of $K$. This process continues until we fail to reject the null hypothesis. Since our proposed bootstrap procedure yields a valid $p$-value, we can directly apply the results in \cite{ghosh2013selecting} to ensure a pre-specified error tolerance. 

Additionally, the proposed \textsf{E2D2} parameter is currently limited to quantifying assortative community structure. Extending the method to handle disassortative and/or bi-partite networks would be a interesting contribution. Next, while the asymptotic test is more adept to scale to large networks, the bootstrap test is limited to networks of up to (roughly) $n=10,000$ nodes for computational feasibility. There are also open theoretical questions including a more precise asymptotic cutoff that accounts for correlation between the random variables as well as bootstrap theory for the maximum test statistic. Moreover, the ideas from this work could be extended to test for other network properties like core-periphery structure \citep{borgatti2000models}.

\bibliographystyle{apalike}
\bibliography{ref}

\section*{Technical Proofs}\label{sec5}

\subsection*{Upper bound on E2D2 parameter}\label{sec1}
We want to show that $\{\bar p_{in}(\bc)-\bar p_{out}(\bc)\}/(K\bar p) \leq 1$. Notice that for any $\bc$, $\bar p= r\bar p_{in}+ (1-r)\bar p_{out}$ for $r=m_{in}/{n\choose 2}$ where $0\leq r\leq 1$ and $r=r(K)$ depends on $K$, the number of communities. Thus, we equivalently want to maximize
\begin{equation}
    f(x,y,r)
    =\frac{x-y}{rx+(1-r)y}
\end{equation}
where $0\leq r,x,y\leq 1$.  First, let's consider a fixed $r$. Then $f(x,y,r)$ will clearly be maximized when $y=0$ which yields
\begin{equation}
    f(x,0,r)
    =\frac1r.
\end{equation}
Thus, $f(x,y,r)$ is maximized when $r$ is minimized, or, equivalently, when $m_{in}$ is minimized for a fixed $K$.

Let $m_k$ be the number of nodes in community $k\in\{1,\dots,K\}$ where $m_1+\cdots+m_K=n$.
Then we want to minimize $m_{in}=\frac12\sum_{j=1}^K m_j(m_j-1)$ subject to $\sum_{j=1}^K m_j=n$. We can use Lagrange multipliers:
\begin{align}
    \mathcal L(m_1,\dots,m_K,\lambda)
    =\frac12\sum_{j=1}^K m_j(m_j-1) - \lambda \left(\sum_{j=1}^K m_j-n\right)
\end{align}
Take the gradient:
\begin{equation}
    \nabla \mathcal L(m_j,\lambda)
    =\left(m_1-\tfrac12-\lambda,\dots,m_K-\tfrac12-\lambda,n-\sum_{j=1}^K m_j \right)
\end{equation}
Setting equal to 0 means that for all $j$,
$
    m_j=\lambda+\frac12
$
so 
\begin{equation}
    0
    =n-\sum_{j=1}^K (\lambda+\tfrac12)
    \implies \lambda = \frac{n}{K}-\frac12.
\end{equation}
Thus, $m_{in}$ is minimized at $m_1=\cdots=m_k=\frac nK$ so
\begin{equation}
    m_{in} \geq \frac12\sum_{j=1}^K \tfrac{n}{K}(\tfrac{n}{K}-1)
    =\frac{n(n-K)}{2K}.
\end{equation}
Thus, 
\begin{equation}
    f(x,y,r)
    \leq \frac1r
    \leq\frac{{n\choose 2}}{n(n-K)/2K}
    =K\frac{n-1}{n-K}.
\end{equation}
For large $n$, $(n-1)/(n-K)\approx 1$ so we have the desired result.

\subsection*{Theorem 2.1}

First, note that since we assume $K_n$ is known, we can ignore it during the proof and simply divide the final cutoff by $K_n$. Now, let $\gamma_0=\xi_0/\bar p$. Assume a rejection region of the form $R=\{T_*(n)>c(n)\}$ where $c(n)=\frac{\xi_0+k(n)}{\bar p(n)/(1+\epsilon)}$ and
\begin{equation}
    T_*(n)
    =\frac{U_*(n)}{S(n)}
\end{equation}
where $U_*(n)=\max_{\bc}\{\hat p_{in}(\bc)-\hat p_{out}(\bc)\}$ with the max taken over all possible community assignments $\bc_i$ for $i=1,\dots,N_{n,K}$; $S(n)=\hat p(n)$ and
$$
    \bar p(n)
    =\frac1{\binom{n}{2}}\sum_{i>j} P_{ij}(n).
$$
From this point, we suppress the dependence on $n$. Using DeMorgan's Law, we can show that
\begin{equation}
    P(T_*>c)
    \leq P(U_*>\xi_0+k) + P(S<\bar p/(1+\epsilon)).
\end{equation}
where
\begin{equation}
    \bar p
    =\frac1{{n\choose 2}}\sum_{i<j}P_{ij}.
\end{equation}
Under $H_0$, we show that each term on the right-hand side goes to 0. Assume the null model $P_0$ and consider a fixed community assignment with $K_n$ communities, $\bc_i$, for $i\in\{1,\dots,N_{n,k}\}$ where $N_{n,K_n}\leq K_n^n$ and let $U_i=\hat p_{in}(\bc_i)-\hat p_{out}(\bc_i)$. Then
\begin{equation}
    U_i
    =\sum_{j<k} X_{jk}
\end{equation}
where $X_{jk}=m_{in,i}^{-1}$ if $(\bc_i)_j=(\bc_i)_k$ and $-m_{out,i}^{-1}$ otherwise. From the proof of the upper bound on the E2D2 parameter, we have that $m_{in,i}=O(n^2)$ and $m_{out,i}=O(n^2)$. Thus, letting $k_i'=\mathsf{E}(U_i)+k$ and using Hoeffding's inequality,
\begin{align}
    \frac{\eta}{N_{n,K}}
    &=P(U_i\geq k_i')\\
    &=P(U_i\geq \mathsf{E}(U_i)+k)\\
    &\leq \exp\left(\frac{-2k^2}{{n\choose 2}(\frac1{m_{in}}+\frac1{m_{out}})^2}\right)\\
    &\leq \exp\left(-n^2k^2\right)\\
    \implies k&\leq \left(\frac{\log N_{n,K}-\log\eta}{n^2}\right)^{1/2}
    \sim \left(\frac{\log K_n}{n}\right)^{1/2}
\end{align}

Now, under the null hypothesis, $\mathsf{E}(U_i)\leq \xi_0$.  Then we have
\begin{multline}
    P\{U_*>\xi_0 + k\}
    =P\left(\bigcup_{i=1}^{N_{n,K}}\{U_i>\xi_0 + k\}\right)\\
    \leq P\left(\bigcup_{i=1}^{N_{n,K}}\{U_i>k'_i\}\right)
    \leq \sum_{i=1}^{N_{n,K}}
    P\{U_i>k'_i\}
    \leq\sum_{i=1}^{N_{n,K}}\frac{\eta}{N_{n,K}}
    \leq\eta.
\end{multline}
We also have
\begin{equation}
    P(S<\bar p/(1+\epsilon))
    =P(S < \bar p -\tfrac{\epsilon}{1+\epsilon}\bar p)
    \leq e^{-\epsilon^2\bar p^2n(n-1)/(1+\epsilon)^2}
    \to 0
\end{equation}
since $n^{1/2}\bar p\to\infty$. Combining these two results we have that 
\begin{equation}
    P(T_*>c)
    \leq P(U_*>\xi_0 + k) + P(S<\bar p/(1+\epsilon))
    \leq \eta
\end{equation}
as we hoped to show.\\ \\

Under $H_1$, let $\gamma_1=\xi_1/\bar p$ and let $T_{oracle}=T(\bc_\gamma,A)=U_{oracle}/S$ where $\bc_\gamma=\arg\max_{\bc}\{\gamma(\bc,P)\}$, i.e., $\bc_\gamma$ is the community assignment which maximizes the E2D2 parameter. This is reasonable because we assume that the algorithm finds the global maximum $\tilde T(A)$ so $T_{oracle}\leq \tilde T(A)$. We will use a similar approach to the proof of $H_0$ noting that
\begin{equation}
    \{U_{oracle}>(\xi_0+k)\tfrac{1+\epsilon}{1-\epsilon}\}\cap \{S\leq \tfrac{\bar p}{1-\epsilon}\}\subseteq \{T_{oracle}>c\}
\end{equation}
so
\begin{align}
    P(T_{oracle}>c)
    &\geq P\{U_{oracle}>(\xi_0+k)\tfrac{1+\epsilon}{1-\epsilon}\}\cap \{S\leq \tfrac{\bar p}{1-\epsilon}\})\\
    &\geq P \{U_{oracle}>(\xi_0+k)\tfrac{1+\epsilon}{1-\epsilon}\} + P\{S\leq \tfrac{\bar p}{1-\epsilon}\} - 1.
\end{align}
Thus, we want to show that the first two terms on the right-side go to 1. For the first term, we note that $U_{oracle}$ is the sum of $O(n^2)$ independent random variables, each of which takes values between $[-m^{-1}_{out}, m_{in}^{-1}]$. Moreover, $\mathsf{E}(U_{oracle})=\xi_1>\xi_0$. Let $1_\epsilon:=(1+\epsilon)/(1-\epsilon)$. Then,
\begin{align}
    P\{U_{oracle} \leq (\xi_0+k)1_\epsilon\}
    &=P\{U_{oracle} \leq \xi_1 1_\epsilon -( \xi_1-\xi_0-k)1_\epsilon\}\\
    &=P\{U_{oracle}\leq \xi_1 - \underbrace{(\xi_1-\xi_0-\tfrac{2\epsilon}{1+\epsilon}\xi_1-k )}_{z}\}.
\end{align}
Now, $z>0$ since $\xi_1-\xi_0>0$ and we can choose $\epsilon$ small enough such that $\xi_1-\xi_0-\frac{2\epsilon}{1+\epsilon}\xi_1>0$. Additionally, $k\to0$ by A3 so there exists an $N$ such that for all $n\geq N$, $\xi_1-\xi_0-\frac{2\epsilon}{1+\epsilon}\xi_1 > k$. Thus, we can use Hoeffding's inequality to show
\begin{align}
    P\{U_{oracle}\leq (\xi_0 + k)1_\epsilon\}
    &=P\{U_{oracle} \leq \xi_1-z\}\\
    &\leq \exp\left(-\frac{2z^2}{\sum_{i=1}^{n^2}\frac1{n^4}}\right)\\
    &= \exp\left(-2n^2z^2\right)\\
    &\to 0,
\end{align}
or equivalently,
\begin{equation}
    P\{U_{oracle} > (\xi_0 +k)1_\epsilon \}
    \to 1.
\end{equation}
Next, consider $S$. First, notice that
\begin{equation}
    \frac{\bar p}{1-\varepsilon}
    =\bar p + \frac{\varepsilon}{1-\varepsilon}\bar p
\end{equation}
Then, by Hoeffding's inequality, we can show 
\begin{align}
    P(S\geq \bar p/(1-\varepsilon))
    &=P(S \geq \bar p + \tfrac{\varepsilon}{1+\varepsilon}\bar p)\\
    &\leq e^{-\varepsilon^2 (\bar p)^2/(1+\varepsilon)^2n(n-1)}\\
    &\to 0
\end{align}
since $n^{1/2}\bar p\to\infty$. Then
\begin{equation}
    \lim_{n\to\infty}P(\tilde T(A) > C)
    \geq \lim_{n\to\infty}P(T_{oracle}>C)
    \geq 1+1-1
    \geq 1. \square
\end{equation}

\subsection*{Proposition in Section 2.4}
{\it Claim: $\tilde\gamma(P)=0$ if and only if $P$ is from an ER model.}

{\it Proof.} The only if direction of the claim is immediate. To prove the forward direction, we first show that $\gamma(\bc,P)\leq0$ for all $\bc$ implies that $\gamma(\bc,P)=0$ for all $\bc$. Then we show that if $\gamma(\bc,P)=0$ for all $\bc$, then $P$ is from an ER model which is equivalent to showing $\tilde\gamma(P)=0$.

For the first part, this is equivalent to showing that if $\gamma(\bc,P)<0$ for some $\bc$, then there exists some $\bc'$ such that $\gamma(\bc',P)>0$. If there exists some $\bc$ such that $\gamma(\bc,P)<0$, then
$$
    \frac1{\sum_{k=1}^K \binom{n_k}{2}}\sum_{i<j}^K\delta_{c_i,c_j}P_{ij}< \frac1{\sum_{k> l} n_k n_l}\sum_{i<j}(1-\delta_{c_i,c_j}) P_{ij}.
$$

But this means that there is some $P_{ij}$ such that $P_{ij}\geq P_{kl}$ for all $i\neq k$ or $j\neq l$ and is strictly greater for at least one $P_{kl}$. Thus, if we consider the community assignment $\bc'$ where nodes $i$ and $j$ are in one community and all other nodes are in the other, then $\bar p_{in}(\bc')>\bar p_{out}(\bc')$ and thus $\gamma(\bc',P)>0$.

We will prove the second part by induction. Let $n=3$ and we are given that $\gamma(\bc,P)=0$ for all $\bc$. We start by writing out the probability matrix.
$$
    P
    =
    \begin{pmatrix}
        - & P_{12} & P_{13}\\
          & - &P_{23}\\
          & & -
    \end{pmatrix}.
$$
There are three possible community assignments: $\bc_1=\{1,1,2\},\bc_2=\{1,2,1\}$ and $\bc_3=\{2,1,1\}$. From each of these assignments, we have a corresponding statement relating the probabilities:
\begin{align*}
    \bar p_{in}=P_{12}&=\bar p_{out}= \frac12(P_{13}+P_{23})\\
    \bar p_{in}=P_{13}&=\bar p_{out}= \frac12(P_{12}+P_{23})\\
    \bar p_{in}=P_{23}&=\bar p_{out} =\frac12(P_{12}+P_{13}).
\end{align*}
Plugging the first equation into the second equation we find:
$$
    P_{13}
    =\frac12(\tfrac12(P_{13}+P_{23})+P_{23})
    \implies 
    P_{13}=P_{23}.
$$
Plugging this into the first equation we have $P_{12}=P_{13}=P_{23}:=p$ which means that this must be an ER model. 

Now assume that the claim holds for $n-1$ and show it holds for $n$. For convenience, assume $n$ is even but the proof can easily be extended if $n$ is odd. Consider a network with $n$ nodes such that $\gamma(\bc,P)=0$. Remove an arbitrary node such that we have a network with $n-1$ nodes and apply the induction hypothesis, i.e. $P_{ij}=p$ for all $i,j$. We now add the removed node back to the network such that the node has probability $P_{i,n}$ of an edge between itself and node $i$ for $i=1,\dots,n-1$. Thus, the probability matrix is:
$$
    P
    =
    \begin{pmatrix}
        -&p&p&\cdots&p&P_{1n}\\
         &-&p&\cdots&p&P_{2n}\\
         & & & \ddots & &\vdots\\
         & &&&&P_{n-1,n}\\
         & &\ddots&&&-
    \end{pmatrix}.
$$
Since $\gamma(\bc,P)=0$ for all $\bc$, then we want to show that $P_{i,n}=p$ for $i=1,\dots,n$. Assume for contradiction that $P$ is not ER and we will show that $\gamma(\bc,P)\neq0$ for some $\bc$. Without loss of generality, let $\{P_{1n},\dots,P_{n/2,n}\}$ be the smaller values of the last column and $\{P_{n/2+1,n},\dots,P_{n-1,n}\}$ be the larger values and consider the community assignment where nodes $\{1,\dots,n/2\}$ are in one community and nodes $\{n/2+1,\dots,n\}$ are in the other community. Then
$$
    \bar p_{in}
    =\frac1{2\binom{n/2}{2}}\left( p\cdot (\tfrac n2-1)^2
    +\sum_{i=n/2+1}^{n-1}P_{i,n}\right)
    >
    \bar p_{out}
    =\frac1{n^2/4}\left(p\cdot(\tfrac{n^2}{4}-\tfrac n2) + \sum_{i=1}^{n/2} P_{i,n}\right)
$$
since
$$
    \sum_{i=n/2+1}^{n-1}P_{i,n}
    >
    \sum_{i=1}^{n/2} P_{i,n}.
$$
Thus $\gamma(\bc,P)\neq0$ for this particular choice of $\bc$ and we have completed the proof. $\square$

\subsection*{Lemma 3.1}
We follow closely the ideas of the proof of Theorem 5 in \cite{levin2019bootstrapping}. Assume that $p\sim F(\cdot)$ and $A,H|p\sim ER(p)$, $\hat A^*|\hat p\sim ER(\hat p)$ where $\hat p=\sum_{i,j}A_{ij}/\{n(n-1)\}$.
We will use the well-known property of Bernoulli random variables that if $X\sim\mathsf{Bernoulli}(p_1)$ and $Y\sim\mathsf{Bernoulli}(p_2)$, then $d_1(X,Y)\leq |p_1-p_2|$. Thus,
$$
    P(\hat A_{ij}^*\neq H_{ij}|p,\hat p)
    \leq |\hat p-p|.
$$
Let $\nu$ be the coupling such that $A$ and $H$ are independent. Then
$$
    W_p^p(\hat A^*, H)
    \leq \int d_{GM}^p (\hat A^*,H) d\nu(\hat A^*,H).
$$
Using Jensen's inequality,
$$
    d_{GM}^p(\hat A^*,H)
    \leq \left(\frac12{n\choose 2}^{-1}||\hat A^*-H||_1\right)^p
    \leq {n\choose 2}^{-1}\sum_{i<j}|\hat A^*_{ij}-H_{ij}|^p
    ={n\choose 2}^{-1}\sum_{i<j}|\hat A^*_{ij}-H_{ij}|.
$$
Thus,
\begin{align*}
    \int d^p_{GM}(\hat A^*,H)d\nu(\hat A^*,H)
    &\leq {n\choose 2}^{-1}\sum_{i<j}\int |\hat A_{ij}^*-H_{ij}|d\nu\\
    &={n\choose 2}^{-1}\sum_{i<j}\nu(\{\hat A_{ij}^*\neq H_{ij}\})\\
    &\leq{n\choose 2}^{-1}\sum_{i<j}|\hat p-p|\\
    &=|\hat p-p|\\
    &=O(n^{-1}). \ \square
\end{align*}

\subsection*{Lemma 3.2}
It's easy to see that the CL model falls into the Random Dot Product Graph framework where $\boldsymbol\theta=(\theta_1,\dots,\theta_n)$ correspond to the latent positions and the dimension $d=1$. Then by Theorem 5 of \cite{levin2019bootstrapping}, we have that
$$
    W_p^p(\hat A^*,H)
    =O((n^{-1/2}+n^{-1/1})\log n)
    =O(n^{-1/2}\log n)
$$
since $\hat{\boldsymbol\theta}$ is estimated using the ASE.

\subsection*{Lemma 3.3}
Let $t(H,\bc)=\sum_{i<j}C_{ij}H_{ij}$ where $C_{ij}=m_{in}^{-1}$ if $c_i=c_j$ and $m_{out}^{-1}$ otherwise and $H_{ij}\sim\mathsf{Bernoulli}(p)$. Define $\mathsf{E}\{t(H,\bc)\}=\xi(H,\bc)$ and
$$
    s_n^2
    =\sum_{i<j} \mathsf{Var}(C_{ij}H_{ij})
    =p(1-p)\sum_{i<j}C_{ij}^2.
$$
We want to invoke Lyapunov's CLT so we must check the follow condition: for some $\delta>0$,
$$
    \lim_{n\to\infty} \frac1{s_n^{2+\delta}}\sum_{i=1}^n \mathsf{E}(|X_i-\mathsf{E}(X_i)|^{2+\delta})\to0.
$$
Let $\delta=1$ and recall that $C_{ij}=O(n^{-2})$. Then, ignoring constants,
\begin{align*}
   \frac1{s_n^3}\sum_{i<j} \mathsf{E}(|C_{ij}H_{ij}-C_{ij}p|^3)
   &=\frac{1}{s_n^3}\sum_{i<j}C_{ij}^3\mathsf{E}(|H_{ij}-p|^3)\\
    &=\frac{1}{s_n^3}\sum_{i<j}C_{ij}^3\\
    &=O(n^3)\sum_{i<j}O(n^{-6})\\
    &=O(n^{-1})\ \checkmark.
\end{align*}
Thus, by Lyapunov's CLT,
\begin{equation}
    \frac1{s_n}\sum_{i<j}(C_{ij}H_{ij}-C_{ij}p)
    =\frac1{s_n}\{t(H,\bc)-\xi(H,\bc)\}\stackrel{d}{\to}\mathsf{N}(0,1).
\end{equation}
Finally, note that $T(H,\bc)=t(H,\bc)/(K\hat p)$ and $\gamma(H,\bc)=\xi(H,\bc)/(Kp)$. Since $\hat p\stackrel{P}{\to}p$, by Slutsky's theorem,
\begin{equation}
    \frac1{s_n}\{T(H,\bc)-\gamma(H,\bc)\}\stackrel{d}{\to}\mathsf{N}(0,K^2p^2).
\end{equation}

\noindent
The results for $\tilde T(\hat A^*,\bc)$ are the same noting that:
$$
    \mathsf{E}(\hat A_{ij}^*)
    =\mathsf{E}(\mathsf{E}(\hat A_{ij}^*|\hat p))
    =\mathsf{E}(\hat p)
    =p
    =\mathsf{E}(H_{ij})
$$  
so $\mathsf{E}(t(\hat A^*,\bc)=\mathsf{E}(t(H,\bc))$;
and
$$
    \mathsf{Var}(\hat A_{ij}^*)
    =\mathsf{Var}(\mathsf{E}(\hat A_{ij}^*|\hat p))
    +\mathsf{E}(\mathsf{Var}(\hat A_{ij}^*|\hat p))
    =\mathsf{Var}(\hat p) +\mathsf{E}(\hat p(1-\hat p))
    =p(1-p)
    =\mathsf{Var}(H_{ij})
$$ 
so $\mathsf{Var}(t(\hat A^*,\bc)=\mathsf{Var}(t(H,\bc))$. $\square$

\end{document}